\documentclass[prl,aps,twocolumn,showpacs,nofootinbib,nobibnotes,superscriptaddress]{revtex4}

\usepackage{epsfig}
\usepackage{amssymb}

\begin{document}


\title{Scaling law of the plasma turbulence with non conservative fluxes}

\date{\today}

\author{Grigol Gogoberidze}
\email{gogober@geo.net.ge} \affiliation{Georgian National
Astrophysical Observatory, 2a Kazbegi ave., 0160 Tbilisi, Georgia}

\begin{abstract}

It is shown that in the presence of anisotropic kinetic dissipation
existence of scale invariant power law spectrum of plasma turbulence
is possible. Obtained scale invariant spectrum is not associated
with the constant flux of any physical quantity. Application of the
model to the high frequency part of the solar wind turbulence is
discussed.

\end{abstract}

\pacs{52.35.Ra, 47.27.Eq, 96.50.Ci}

\maketitle

Dissipation range of incompressible hydrodynamic turbulence was
extensively studied by different authors
\cite{CD93,DR90,YB91,SB05,DSP97,G98,SSY94,SJ93,PBC93} due to the
fact that smallest scale perturbations display strong intermittency
even at Reynolds numbers so low that there is no basis for fractal
cascade. The kinetic energy spectrum $E(k)$ of the hydrodynamic
turbulence in the far dissipation range behaves as
\begin{equation}
E(k) \sim k^{\alpha_1} \exp\left[ - \alpha_2 (k/k_d)^n \right],
\label{eq:00}
\end{equation}
where $k_d$ is Kolmogorov dissipation wave number, $\alpha_1$ and
$\alpha_2$ are some constants and $1 \leq n \leq 2$ (see, e.g.,
\cite{CD93} and references therein).

In the case of plasma turbulence existence of various kinetic
mechanisms of dissipation makes the situation much more complicated.
For instance, observations of the solar wind turbulence
\cite{C68,DBN83,LSNMW98} strongly suggest steep power law spectrum
of the magnetic field fluctuations for the frequencies higher then
ion cyclotron frequency for which kinetic mechanisms of dissipation
are dominant. In contrast with viscosity kinetic dissipation of
plasma waves in the presence of background magnetic field is usually
strongly anisotropic. In the presented paper we show that in the
presence of: (i) anisotropic kinetic dissipation; and (ii) if the
nonlinear transfer is governed by the scattering of the plasma waves
by low frequency ones, then one should expect scale invariant power
law spectrum of the plasma turbulence. It should be emphasized that
the scale invariant spectrum is not associated with the constant
flux of any physical quantity due to the presence of kinetic
dissipation.

The general equation that governs the evolution of any averaged
characteristic $Z$ of the homogenous turbulence which is conserved
by nonlinear interactions (in the case of hydrodynamic turbulence
$Z$ is usually associated with energy density) in the wave number
space has the form
\begin{equation}
\frac{\partial Z}{\partial t} = J + D+S, \label{eq:01}
\end{equation}
where $S$ and $D$ describes the source and dissipation of $Z$ and
$J$ accounts for the accumulation of $Z$ due to the nonlinear
interactions among the various wave number components of the
turbulent field, such as velocity and magnetic fields. If nonlinear
transfer term $J$ serves only to redistribute $Z$ and not changes
total amount, then one can define the flux field $F$ in the wave
number space \cite{L67}
\begin{equation}
J = -\nabla \cdot {\bf F(k)}, \label{eq:02}
\end{equation}
so the property of conservation is automatically fulfilled.

Further we assume that nonlinear interactions are local in the
sense that the most contribution to ${\bf F(k)}$ is from nearby
regions of the ${\bf k}$ space. We make diffusion approximation to
the wave number space transport, i.e., we assume that the flux can
be presented as
\begin{equation}
F_i({\bf k}) = - D_{ij} \frac{\partial Q}{\partial k_j},
\label{eq:03}
\end{equation}
where $D_{ij}$ are diffusion coefficients and $Q$ is potential,
that in general case is some function of ${\bf k}$ and $Z$. So in
the inertial range Eq. (\ref{eq:01}) takes the form
\begin{equation}
\frac{\partial Z}{\partial t} =  \frac{\partial }{\partial k_i}
\left[ D_{ij}(k,Z) \frac{\partial Q(k,Z)}{\partial k_j} \right].
\label{eq:04}
\end{equation}

The diffusion approximation for isotropic hydrodynamic turbulence
was first introduced by Leith \cite{L67}. Afterwards the same
concept was successfully applied to the plasma turbulence both in
the strong \cite{ZM92} and weak \cite{ZK78} turbulence regimes. For
isotropic hydrodynamic turbulence $Z$ corresponds to energy density
${\mathcal E}({\bf k})$, $D_{ij} \sim k^{9/2} \delta_{ij}$ and $Q
\sim {\mathcal E}^{3/2}$ \cite{L67}, where $\delta_{ij}$ is
kronecker delta. Combining Eqs. (\ref{eq:01})-(\ref{eq:03}) one can
readily obtain the famous Kolmogorov spectrum $\mathcal E \sim
k^{-11/3}$ for the inertial range of the hydrodynamic turbulence. In
the case of the plasma turbulence the situation is more complicated.
Existence of the background magnetic field usually leads to the
anisotropy of the nonlinear cascade.

Consider plasma turbulence in some frequency range where there is no
source of the turbulence ($S=0$) and there exist kinetic dissipation
of plasma waves. As it was shown in Refs. \cite{DR90, YB91}, energy
transfer in the dissipation range of hydrodynamic turbulence is
dominated by nonlocal triads in which one leg is in the
energy-containing (low-k) range. Similarly, we suppose that the
strongest nonlinear interaction is the scattering of the high
frequency waves by low frequency ones from the inertial range of the
plasma turbulence. In the weak turbulence \cite{K65} theory this
process requires fulfillment of the resonant conditions
\begin{equation}
{\bf k}_1 = {\bf k}_2 +{\bf K}, ~~~\omega_1=\omega_2+\Omega,
\label{eq:04a}
\end{equation}
where ${\bf k}_{1,2}$, $\omega_{1,2}$ and ${\bf K}$, $\Omega$ are
wave numbers and frequencies of high and low frequency waves
respectively. Due to the fact that $\omega \gg \Omega$ and $ k_{1,2}
\gg K$, the change of the wave number is small in the unit act of
the scattering ($|{\bf k}_{1}-{\bf k}_{2}| \ll |{k}_{1,2}|$) and
therefore diffusion approximation is applicable (see, e.g.,
\cite{ZK78,ZLF}).

We incorporate the dissipation in the model as follows: we assume
that the dissipation of the waves is negligible compared to the
nonlinear interaction if the angle of the propagation with respect
to the external magnetic field $\theta$ is less then some angle
$\theta_0$, whereas the opposite limiting case takes place for
$\theta>\theta_0$. This model seems reasonable for the transverse
waves in collisionless plasma, such as whistler waves and
electromagnetic (ordinary and extraordinary) waves in
electron-positron plasma \cite{VKM85,AB86}. Indeed, in this case the
main mechanisms of dissipation are Landau and cyclotron damping.
Both Landau and multiple cyclotron resonances do not affect the
transverse waves for parallel propagation with respect to the
background magnetic field (see, e.g., \cite{ABR}), whereas for
relatively large angles of propagation both mechanisms can be on
work. In the case of Landau damping this is caused by the fact that
ambient propagating wave has nonzero electric field component
parallel to the background magnetic field - necessary condition of
Landau damping.

The second assumption of the presented model, that the nonlinear
transfer is governed by the scattering of plasma waves by low
frequency ones is also satisfied for both kinds of mentioned above
wave modes. For of whistler waves in the solar wind, as it is shown
below the strongest nonlinear process is scattering of whistler
waves by low frequency magnetohydrodynamic waves from inertial range
of the solar wind turbulence. Similarly, the strongest nonlinear
process that governs evolution of electromagnetic waves in electron
positron plasma is their scattering by low frequency Langmuir waves
(see e.g., \cite{MG05} and references therein).

It should be noted that presented model is not valid  for Alfv\'enic
turbulence. Alfv\'en waves have low frequency compared to the ion
cyclotron frequency and therefore they are not affected by cyclotron
damping. Landau damping of Alfv\'en waves also have unusual
properties \cite{M92} incompatible with the presented model.
Additionally, the second assumption about nonlinear transfer is also
violated in this case (see, e.g., \cite{GNNP02}).

Note, that if $\theta_0$ is not extremely small, presented model
implies that the wave with $\theta=0$ should take part in many
scattering events before it can be transferred to the dissipation
area. This circumstance allows us to use Eq. (\ref{eq:04}) for the
conical area in the wave number space $\theta<\theta_0$, and take
the dissipation into account by requesting $Q({\bf k})$ to vanish at
$\theta=\theta_0$.

In the case of Alfv\'enic turbulence numerical simulations
\cite{SMM83} as well as analysis of three and four wave resonant
conditions provides that turbulent cascade is strongly anisotropic.
In the weak turbulent regime there is no cascade in the direction
parallel to the background magnetic field at all \cite{GS97}. When
the dispersion of both high and low frequency waves can be
considered as nearly isotropic for $\theta<\theta_0$, then one can
expect different diffusion coefficients for the directions parallel
and perpendicular to the wave vector ${\bf k}$ of the high frequency
waves. We consider scale invariant diffusion, i.e., assume
$D_{ij}(k,Z)= d_{ij} \delta_{ij} k^{\alpha_i} $, and $Q\sim
k^{\beta_1} Z^{\gamma_1} \equiv k^{\beta_2} E(k)^{\gamma_2}$ where $
d_{ij}$ are not functions of $k$ and $Z$.

With these assumptions Eq. (\ref{eq:04}) reduces to the following
one
\begin{equation}
\frac{1}{k^2} \frac{\rm d}{{\rm d} k} \left( k^2 D_\parallel
\frac{\rm d Q}{{\rm d} k} \right) + \frac{D_\perp}{k^2
\sin\theta}\frac{\rm d}{{\rm d} \theta} \left( \sin \theta
\frac{{\rm d} Q}{{\rm d} \theta}\right)=0, \label{eq:10}
\end{equation}
with boundary condition $Q(k,\theta_0)=0$. Here $D_\parallel =
d_\parallel k^{\alpha_\parallel}$ and $D_\perp = d_\perp
k^{\alpha_\perp} k^{\alpha_\parallel - \alpha_\perp}_s$, and $k_s$
denotes the minimal wave number for which the formulated model is
valid. Let us first consider the case when parallel and
perpendicular diffusion coefficients have the same scaling law,
i.e., $\alpha_\parallel = \alpha_\perp \equiv \alpha$. Using
standard methods of variable separation, i.e., representing
potential as $Q(k,\theta)=R(k) \Psi (\theta)$ we obtain for the
solution of Eq. (\ref{eq:10})
\begin{equation}
Q=\sum_{m=1}^\infty B_m k^{-c_m} P_{\nu_m}(\cos\theta),
\label{eq:11}
\end{equation}
where coefficients $B_m$ are determined by the source of the
turbulence at small wave numbers $k=k_s$, $P_{\nu_m}(\cos\theta)$
are Legendre functions of the first kind and $\nu_m$ are the
solutions of the eigenvalue problem $ P_{\nu}(\cos\theta_0)=0 $,
arranged in order of increasing magnitude and
\begin{equation}
c_m = \frac{1}{2} \left[ -\alpha -1 - \sqrt{(\alpha+1)^2 + 4 \nu_m^2
\frac{d_\perp}{d_\parallel}} \right]. \label{eq:11a}
\end{equation}
If $\theta_0<\pi/2$ the value of the first eigenvalue $\nu_1$ can be
approximated as \cite{AS}
\begin{equation}
\nu_1 \approx \frac{2.405}{\theta_0}-\frac{1}{2}. \label{eq:13}
\end{equation}

For high wave numbers ($k \gg k_s$) the leading term is
\begin{equation}
Q \approx B_1 k^{-c_1} P_{\nu_1}(\cos\theta). \label{eq:14}
\end{equation}
Note that without kinetic dissipation isotropic solutions of Eq.
(\ref{eq:10}) is $Q \sim k^{-\alpha-1}$~[$E(k) \sim
k^{-(\alpha+1+\beta_2)/\gamma_2}$], which correspond to constant
flux of $Z$. In contrary, obtained scale invariant asymptotic
solution [see, Eq. (\ref{eq:14})] is not associated with constant
flux of any physical quantity, due to the presence of kinetic
dissipation. Alternatively, in contrast with hydrodynamic
turbulence, where the spectrum in the dissipation range is
exponential, obtained result shows that the energy spectrum should
decrease as a power law in the dissipation range if the diffusion
coefficients have the same scaling law.

Now consider the case $\alpha_\perp -\alpha_\parallel \equiv
\Delta \alpha \neq 0$. Using the same technique of variable
separation and introducing new variables $K = k/k_s$ and
$P(K)=k^{1+ \alpha_\parallel/2} R(K)$, Eq. (\ref{eq:10}) yields
\begin{equation}
K^2 \frac{\rm d^2 P}{{\rm d} K^2}  - \left[ \frac{\alpha_\parallel
(\alpha_\parallel + 2)}{4} + \nu_m^2 \frac{d_\perp}{d_\parallel}
K^{\Delta \alpha} \right] P =0. \label{eq:14a}
\end{equation}
If $\Delta \alpha <0$, for $K \gg 1$ one can drop the second term
in the squire brackets. This yields the result that coincides with
the result of the isotropic case $Q \sim k^{-\alpha_\parallel
-1}$, i.e., asymptotically dissipation has no influence on the
cascade.

On the other hand, when $\Delta \alpha > 0$, for $K \gg 1$ one can
drop the first term in the squire brackets. Obtained equation can be
solved in the terms of modified Bessel functions. Using asymptotic
properties of modified Bessel functions \cite{AS} we obtain
\begin{equation}
Q \sim K^{-(\alpha_\parallel+1)/2 - \Delta \alpha/4} \exp \left( -
\frac{2\nu_1}{\Delta \alpha_\parallel}
\sqrt{\frac{d_\perp}{d_\parallel}} K^{\Delta \alpha_\parallel/2}
\right), \label{eq:14b}
\end{equation}
Consequently, in the case under consideration the spectrum is
exponential. Although it should be noted, that the decay is more
soft than in the hydrodynamic turbulence when $\Delta \alpha$ is
relatively small ($\Delta \alpha <2$).

One of the possible application of the presented model is high
frequency part of the solar wind turbulence spectrum. Various
spacecraft observations show the presence of persistent magnetic
fluctuations in the solar wind over a broad range of frequencies
\cite{C68,DBN83,LSNMW98}. For low frequencies ($ f \lesssim
10^{-2}-10^{-3}$ Hz) the magnetic field spectrum vary as
approximately $E_M(f) \sim f^{-1}$. For higher frequencies, up to
proton cyclotron frequency ($f\sim 0.1-1$ Hz) the Kolmogorov
spectrum $f^{-5/3}$ is observed. This is believed to be the inertial
interval of the solar wind turbulence. The change of slope and rapid
decrease in the intensity near the ion cyclotron frequency is
usually considered to be due to the absorption of Alfv\'en waves by
ion cyclotron damping or Landau damping \cite{LSNW99}. At the
frequencies, higher then ion cyclotron frequency, weak but
persistent level of magnetic fluctuations, that can be well
approximated by the power law spectrum  $f^{-3}$ is observed up to
the electron cyclotron frequency. These fluctuations are usually
associated with the whistler waves \cite{DBN83}. The nature of this
high frequency part of the spectrum remains unexplained.

Whistler turbulence have been intensively studied by different
authors both in strong \cite{BSD96,CL04,KM05} and weak
\cite{TS77,GB03} turbulent regimes. If one assumes the existence of
the inertial interval of the whistler turbulence, then
Kolmogorov-type dimensional analysis yields for the magnetic
spectrum \cite{BSD96} $E_M(k) \sim k^{-7/3}$, that is incompatible
with observations (note that due to the relation $E_M(f)df \sim
E_M(k)dk$, and taking into account Doppler shift and dispersion of
whistler waves $f\sim k^2$, observed $f^{-3}$ spectrum corresponds
to $k^{-v}$ with $v \sim 5-6$ in the wave number space).

There exist several different directions of the research for
explanation of high frequency solar wind spectrum. The first
approach \cite{GSRF96,KM05} is based on the fact that governing
equations of Hall magnetohydrodynamics besides energy, conserves two
other second order (with respect to the field variables) quantities
- magnetic and generalized helicity \cite{MY98}. Therefore,
stationary Kolmogorov-type spectrum can be "driven" not only by
energy cascade, but also by the cascade of magnetic and generalized
helicities \cite{KM05}. In Ref. \cite{SGL01} short wavelength
dispersive properties of the magnetosonic-whistler waves have been
studied as a possibility reason of the spectrum steepening.
Alternative approach to the explanation of the high frequency
magnetic fluctuations spectrum in the solar wind implies
incorporation of the linear kinetic effects, such as Landau and
cyclotron damping. It has been shown \cite{LGS01} that simple
incorporation of dissipation term to the energy budget equation
leads to sharp cut off of the energy spectrum. On the other hand,
total ignore of dissipation leads to much more smooth spectrum
compared to the observed one.

The model considered in the presented paper could have important
consequences for the explanation of the high frequency part of the
solar wind spectrum. Whistler waves propagating along the background
magnetic field are affected by neither Landau nor cyclotron damping
on multiple ion cyclotron resonances. Based on the numerical
solution of linear Vlasov equation \cite{LGS01} the angle $\theta_0$
at which kinetic dissipation becomes dominant can be estimated as
$\theta_0 \sim \pi/6$. The level of whistler wave fluctuations is
low in the sense that $\langle b^2_w \rangle/B_0^2 \ll 1$, where
$\langle b^2_w \rangle$ is rms of the whistler wave magnetic field
fluctuations and ${\bf B}_0$ is the background magnetic field.
Therefore the study can be held in the framework of the weak
turbulence theory. Possible nonlinear processes includes: (a) four
wave resonant interactions of whistler waves [it can be shown that
if $\theta_0<\pi/3$ then three wave resonances of whistler waves are
absent, i.e., with this restriction for all three waves resonant
conditions similar to (\ref{eq:04a}) do not have nontrivial
solutions]; (b) induced scattering of whistler waves by ions; and
(c) scattering of whistler waves by low frequency
magnetohydrodynamic waves from the inertial range of the turbulence,
i.e., three wave interactions which involves two whistlers and one
magnetohydrodynamic wave. Detailed analysis of the nonlinear
processes of the solar wind whistler waves will be presented
elsewhere. Here we note that characteristic time scales of these
processes are respectively proportional to $\tau_a \sim N^{-2}$,
$\tau_b \sim N^{-1}$, and $\tau_c \sim 1$, where $N({\bf k}) \equiv
{\mathcal E({\bf k})/\omega({\bf k})}$ is the number density of
whistler waves. Consequently, the strongest nonlinear process that
should be responsible for the formation of the high frequency
spectrum is the scattering of whistler waves by low frequency
magnetohydrodynamic ones. This process conserves the total number of
whistler waves \cite{ZK78}, and therefore $Z \equiv N(\bf k)$.

It can be shown that Alfv\'en waves do not interact with whistlers
through three wave resonances, whereas kinetic Alfv\'en waves do
\cite{C77}. Another possibility is scattering of whistler waves by
fast magnetosonic waves. Here we consider only the second
possibility.

Analytical calculations of diffusion coefficients are very
complicated even in the incompressible limit. In the presented paper
we perform qualitative analysis of three wave interaction of
whistler and fast magnetosonic waves and determine relations between
diffusion coefficients that correspond to the observed spectrum. For
this purposes we use equations of incompressible Hall
magnetohydrodynamics \cite{MY98}
\begin{equation}
\frac{\partial {\bf B}}{\partial t} = {\bf \nabla} \times \left[
\left( {\bf V} - {\bf \nabla} \times {\bf B}\right) \times {\bf
B}\right], \label{eq:15}
\end{equation}
\begin{equation}
\frac{\partial ({\bf B}+ {\bf \nabla} \times {\bf V})}{\partial t} =
- {\bf \nabla} \times \left[ \left( {\bf B} - {\bf \nabla} \times
{\bf V}\right) \times {\bf V}\right], \label{eq:16}
\end{equation}
where time and space variables are measured in units of ion
giroperiod $\omega_{ic}^{-1}$ and ion skin depth $\lambda_i$,
respectively. Analysis of these equations shows that for triad
interaction of whistler and fast magnetosonic waves the strongest
nonlinear term is the second one on the right side of Eq.
(\ref{eq:15}). Taking this into account and noting that the wave
number of whistler waves is much greater then the wave number of
fast magnetosonic waves, then Fourier transform of Eq. (\ref{eq:15})
yields
\begin{equation}
\frac{\partial {\bf b}_{\bf k}^w}{\partial t} - k_{\parallel} B_0
({\bf k} \times {\bf b}_{\bf k}^w)= i {\bf k} \times \left[ {\bf
b}^f \times \left( {\bf \nabla} \times {\bf b}^w \right)
\right]_{\bf k}, \label{eq:17}
\end{equation}
where index ${\bf k}$ denoted the Fourier transform and superscripts
$w$ and $f$ indicates that corresponding values correspond to
whistler and fast magnetosonic waves, respectively.

Using helicity decomposition \cite{GB03} one can apply the standard
technique of weak turbulence theory developed for two types of
interacting waves \cite{ZK78}. But we do not perform this analysis
here due to the reason that only thing that we need for further
analysis is scaling index of the matrix element of interaction $T$
that immediately follows from Eq. (\ref{eq:17}): $T \sim k^2$. In
the framework of weak turbulence theory the dynamics is totally
determined by linear dispersions of waves and matrix element of
interaction. The method of finding the scaling index of diffusion
coefficient was developed in Ref. \cite{ZK78}, which in the case
under consideration yields $\alpha = -1$ and
$\beta_1=0,~\gamma_1=1$. Therefore, Eq. (\ref{eq:14}) yields
\begin{equation}
N({\bf k}) \sim k^{-c_1}, \label{eq:18}
\end{equation}
where $c_1=\nu_1 (d_\perp /d_\parallel)^{1/2}$.

To obtain spectral index $\delta$ of corresponding energy spectrum
$E(f) \sim f^{-\delta}$, we note that
\begin{equation}
E(f) \sim E(k) \frac{{\rm d}k}{{\rm d} f} \sim  {\mathcal E}({\bf
k}) k^2 \frac{{\rm d}k}{{\rm d} f} \sim N({\bf k}) k^3.
\label{eq:19}
\end{equation}
Taking also into account that for whistler waves $f \sim k^2$, Eqs.
(\ref{eq:18}) and (\ref{eq:19}) yield
\begin{equation}
\delta = \frac{c_1-3}{2}. \label{eq:20}
\end{equation}
As it was mentioned above, according to observations $\delta \approx
3$. Taking also $\theta_0 = \pi/6$, Eqs. (\ref{eq:13}) and
(\ref{eq:20}) yield $d_\perp/d_\parallel \approx 5$. Obtained result
seems reasonable, due to the fact that in the magnetized media
perpendicular cascade rate usually significantly exceeds parallel
cascade rate (see, e.g., \cite{CB03} and references therein).

In the presented paper plasma turbulence in the presence of
anisotropic kinetic dissipation is considered. It is shown that if
the nonlinear transfer is governed by the scattering of the plasma
waves by low frequency ones, then development of asymptotic scale
invariant power law spectrum of the plasma turbulence is possible.
Obtained scale invariant spectrum is not associated with the
constant flux of any physical quantity due to the presence of
kinetic dissipation. Corresponding spectral index is given by Eq.
(\ref{eq:11a}) with $m=1$. Possible application of the presented
model to the high frequency part of the solar wind spectrum has been
analyzed.

Author is immensely grateful to Alexander Bershadskii and George
Machabeli for helpful discussions and suggestions.



\end{document}